\documentclass{PoS}

\newcommand\Eq[1]{Eq.~\ref{eq:#1}}
\newcommand\Fig[1]{Fig.~\ref{fig:#1}}
\newcommand\Sec[1]{Sec.~\ref{sec:#1}}
\newcommand\bfn{\mathbf n}
\newcommand\bfe{\mathbf e}
\newcommand\bfp{\mathbf p}
\newcommand\bfq{\mathbf q}

\newcommand\calO{\mathcal O}
\newcommand\calT{\mathcal T}
\newcommand\calV{\mathcal V}
\newcommand\calD{\mathcal D}
\newcommand\calH{\mathcal H}

\title{A new approach for studying large numbers of fermions in the unitary regime} 
\ShortTitle{A new approach for studying large numbers of fermions in the unitary regime} 

\author{
\speaker{Michael G. Endres}\\
Physics Department, Columbia University, New York, NY 10027, USA and\\ 
Theoretical Physics Laboratory, RIKEN, Wako, Saitama 351-0198, Japan\\ 
E-mail: \email{endres@riken.jp}}

\author{
David B. Kaplan\\
Institute for Nuclear Theory, University of Washington, Seattle, WA 98195-1550, USA\\
E-mail: \email{dbkaplan@phys.washington.edu}
}

\author{%
Jong-Wan Lee\\
Institute for Nuclear Theory, University of Washington, Seattle, WA 98195-1550, USA\\
E-mail: \email{jwlee823@u.washington.edu}
}

\author{%
Amy N. Nicholson\\
Institute for Nuclear Theory, University of Washington, Seattle, WA 98195-1550, USA\\
E-mail: \email{amynn@u.washington.edu}
}

\abstract{
A novel lattice approach is presented for studying systems comprising a large number of interacting nonrelativistic fermions.
The construction is ideally suited for numerical study of fermions near unitarity--a strongly coupled regime corresponding to the two-particle s-wave scattering phase shift $\delta_0=\pi/2$.
Such systems may be achieved experimentally with trapped atoms, and provide a starting point for an effective field theory description of nuclear physics.
We discuss the construction of our lattice theory, which allows us to study systems of up to (but by no means limited to) 38 fermions with high accuracy and modest computational resources, and offer an overview of several applications of the technique.
A more detailed discussion of applications and simulation results will be described in companion proceedings by A. N. N. and J-W. L.
}

\FullConference{
The XXVIII International Symposium on Lattice Field Theory, Lattice2010\\
June 14-19, 2010\\
Villasimius, Italy}

\begin{document}

\section{Introduction}
\label{sec:1}

Simulating theories at finite fermion density has been a long-standing challenge in lattice field theory.
One of the main issues faced in Monte Carlo studies of such theories is the fermion sign problem: when a chemical potential is introduced, the action often becomes complex.
A consequence of this is that the exponential of the action cannot be interpreted as a probability measure as required by standard Monte Carlo algorithms.
In the case of lattice QCD at finite baryon number density, the sign problem has severely limited the explorable regions of the QCD phase diagram in the temperature-chemical potential ($T$-$\mu_B$) plane to the regime where $\mu_B/T \lesssim 1$.

Alternatively, at zero temperature, finite densities can be achieved in a canonical ensemble setting by considering multi-fermion correlation functions in a finite box.
In this case, however, a closely related problem emerges: correlators involving fermions typically have a signal/noise which decays exponentially with the time separation of the source and sink.
This makes identification of effective mass plateaus difficult--if not impossible--for large numbers of fermions.
In QCD, standard Lepage arguments \cite{Lepage:1989hd} suggest that the signal/noise for multi-baryon correlation functions decay with a time constant $\tau_B^{-1}\sim B (m_p - 3/2 m_\pi)$, where $B$ is the baryon number, $m_p$ the proton mass and $m_\pi$ the pion mass.

In an effort to better understand the sign and signal/noise problems as well as their interrelationship, we choose to study multi-fermion systems in a much simpler setting than QCD.
One of the simplest nontrivial and interesting theories describes a dilute two-component system of nonrelativistic fermions in the unitary regime.
This regime corresponds to a two particle s-wave scattering phase shift near $\delta_0 = \pi/2$, and when this limit is achieved, the bound $\sigma_{l=0} \leq 4\pi/p^2$ on the s-wave scattering cross-section becomes saturated.
Although these systems are dilute, they are strongly interacting and therefore require a nonperturbative treatment for reliable study.
Unitary fermions are interesting not only because they can be studied experimentally using trapped ultra-cold atoms, but also because they serve as a starting point for a lattice effective field theory (EFT) description of nuclear physics.
The latter is due to the fact that the ${}^1S_0$ and ${}^3S_1$ scattering lengths for nucleon-nucleon scattering are unnaturally large compared to the range of interaction.

In this work, we focus on a new canonical ensemble approach for simulating large numbers of nonrelativistic fermions in the unitary regime and at zero temperature.
Using this new approach, we consider two systems in particular: unitary fermions in a finite box and unitary fermions in a harmonic trap.
From numerical simulations of these systems, one may extract experimentally measurable non-perturbative quantities such as the Bertsch parameter (i.e., the ratio of the energy to that of the free gas energy in the thermodynamic limit) and pairing gap.
Furthermore, moving away from unitarity one may test a set of universal relations involving a quantity known as the ``integrated contact density'', first discovered by Tan \cite{2008AnPhy.323.2952T,2008AnPhy.323.2971T,2008AnPhy.323.2987T}.

These proceedings focus primarily on the details of our lattice construction, as well as the simulation and parameter tuning methods used in our studies of unitary fermions.
Many past numerical simulations at zero temperature have been variational in nature, offering only an upper bound on the Bertsch parameter and possessing unknown systematic errors on other quantities \cite{Chang:2004zz,Chang:2007zz,2010arXiv1008.3191B}.
Our approach is nonvariational and is therefore in principle free from such systematic errors.
However, as is common among all simulations of this type, obtaining reliable results for the spectrum and matrix elements requires a good choice of interpolating operators which possess large overlap with the states of interest.
Details of how we construct optimal sources and sinks as well as consideration of systematic errors due to finite volume and lattice spacing effects are the focus of our companion proceedings \cite{Lee:2010qp,Nicholson:2010ms}.
In those proceedings we also present results of our initial exploratory studies of up to 20 trapped and 38 untrapped fermions, and present preliminary values for the Bertch parameter and pairing gap.

\section{Lattice construction}
\label{sec:2}

The starting point for our construction is a highly improved variant of the nonrelativistic lattice action first proposed in \cite{Chen:2003vy}, given by:
\begin{eqnarray}
S = \sum_\bfn \left[ \bar\psi_\bfn  (\partial_\tau \psi)_\bfn  - \frac{1}{2M} \bar\psi_\bfn (\nabla^2 \psi)_\bfn + \phi_\bfn \bar\psi_\bfn (\sqrt{C} \psi)_{\bfn - \bfe_0} \right]\ .
\label{eq:action}
\end{eqnarray}
This action describes two species of interacting fermions $\psi = (\psi_{\uparrow}, \psi_{\downarrow})$ defined on a $T\times L^3$ lattice with open boundary conditions in the time direction and periodic boundary conditions in the space directions.
The derivative $\partial_\tau$ represents a forward difference operator in time and $\nabla^2$ is a non-local lattice gradient operator which we define so-as to give a ``perfect'' continuum-like single particle dispersion relation for free fermions.
A four-fermion contact interaction is achieved via a Gaussian or $Z_2$ auxiliary field $\phi$ associated with the time-like links of the lattice.
The operator $C$ acts only in space, and in principle may include derivative interactions to an arbitrary even order in momenta.

We may express \Eq{action} succinctly as $S=\bar\psi K \psi$, where the time components of the fermion matrix $K$ are given in block-matrix form by:
\begin{eqnarray}
K = \left( \begin{array}{cccccc}
D      & X(0)   & 0      & 0      & \ldots & 0      \\
0      & D      & X(1)   & 0      & \ldots & 0      \\
0      & 0      & D      & X(2)   & \ldots & 0      \\
0      & 0      & 0      & D      & \ldots & 0      \\
\vdots & \vdots & \vdots & \vdots & \ddots & X(T-1) \\
0      & 0      & 0      & 0      & \ldots & D      \\
\end{array} \right)\ ,
\end{eqnarray}
with
\begin{eqnarray}
D = 1 - \frac{\nabla^2}{2M} \ ,\qquad X(\tau) = 1 - \phi(\tau) \sqrt{C} \ .
\end{eqnarray}
Note that the $L^3 \times L^3$ matrices $D$, $X$ and $C$ act only in space and that $\phi(\tau)$ is a diagonal matrix with independent random elements.
In momentum space the specific expressions we use for $D$ and $C$ are:
\begin{eqnarray}
\langle \bfp | D | \bfp' \rangle = \left\{ \begin{array}{ll}
e^{\bfp^2/(2M)} \delta_{\bfp,\bfp'}& |\bfp| < \Lambda \\
\infty  & |\bfp| \ge \Lambda
\end{array} \right. \ ,\qquad
\langle \bfp | C | \bfp' \rangle = C(\bfp) \delta_{\bfp,\bfp'}
\label{eq:d}
\end{eqnarray}
where $\Lambda=\pi$ is a hard momentum cutoff imposed on the fermions and $C(\bfp)$ is some analytic function of $\bfp^2$ which may be determined order by order in momenta from scattering data (see \Sec{5} for details).

Fermion propagators from time slice zero to time slices $\tau$ may be expressed as a sequence of applications of $D$ and $X$ operators:
\begin{eqnarray}
K^{-1}(\tau;0) = D^{-1} X(\tau-1)  D^{-1} \ldots  D^{-1} X(0)  D^{-1}\ ,
\label{eq:prop}
\end{eqnarray}
which provides a simple recursive approach for computation.
Inversion of the non-local $D$ operator and application of $X(\tau)$ may be performed efficiently with fast Fourier transforms (FFTs); it is this feature that allows us to use a perfect dispersion relation and momentum dependent interactions.
In the free-field limit, one may explicitly verify that our definition for $D$ provided in \Eq{d} yields a perfect dispersion relation for fermions.
In momentum space the fermion propagator reduces to:
\begin{eqnarray}
\langle \bfp | K^{-1}(\tau;0) | \bfp' \rangle = e^{- E(\bfp) \tau} \delta_{\bfp,\bfp'} \ ,\qquad E(\bfp) = \frac{\bfp^2}{2M}\ ,
\end{eqnarray}
for momenta less than the cut-off $\Lambda$.

\section{Simulation method and cost}
\label{sec:3}

Since $K$ is an upper tri-diagonal block matrix, the fermion determinant obtained by ``integrating out'' the fermions is given by $\det K = \det D^T$. 
We see that the determinant is independent of the auxiliary field, and therefore the full numerical simulation of \Eq{action} is equivalent to a quenched simulation.
By construction, our simulations are therefore free of the sign problem.
Furthermore, the decomposition of fermion propagators as a product of random matrices in \Eq{prop} provides an unusual interpretation for our approach: in essence we perform Euclidean time-evolution of single particle wave functions over random background noise (either $Z_2$ or Gaussian).
Multi-particle sources and sinks may be constructed from a direct product of single particle wavefunctions, and in the case of sinks, more elaborate constructions may be considered as well which incorporate pairing correlations \cite{Lee:2010qp, Nicholson:2010ms}.

\begin{figure}
\begin{minipage}[t]{0.486\textwidth}
\centering
\includegraphics[width=\textwidth]{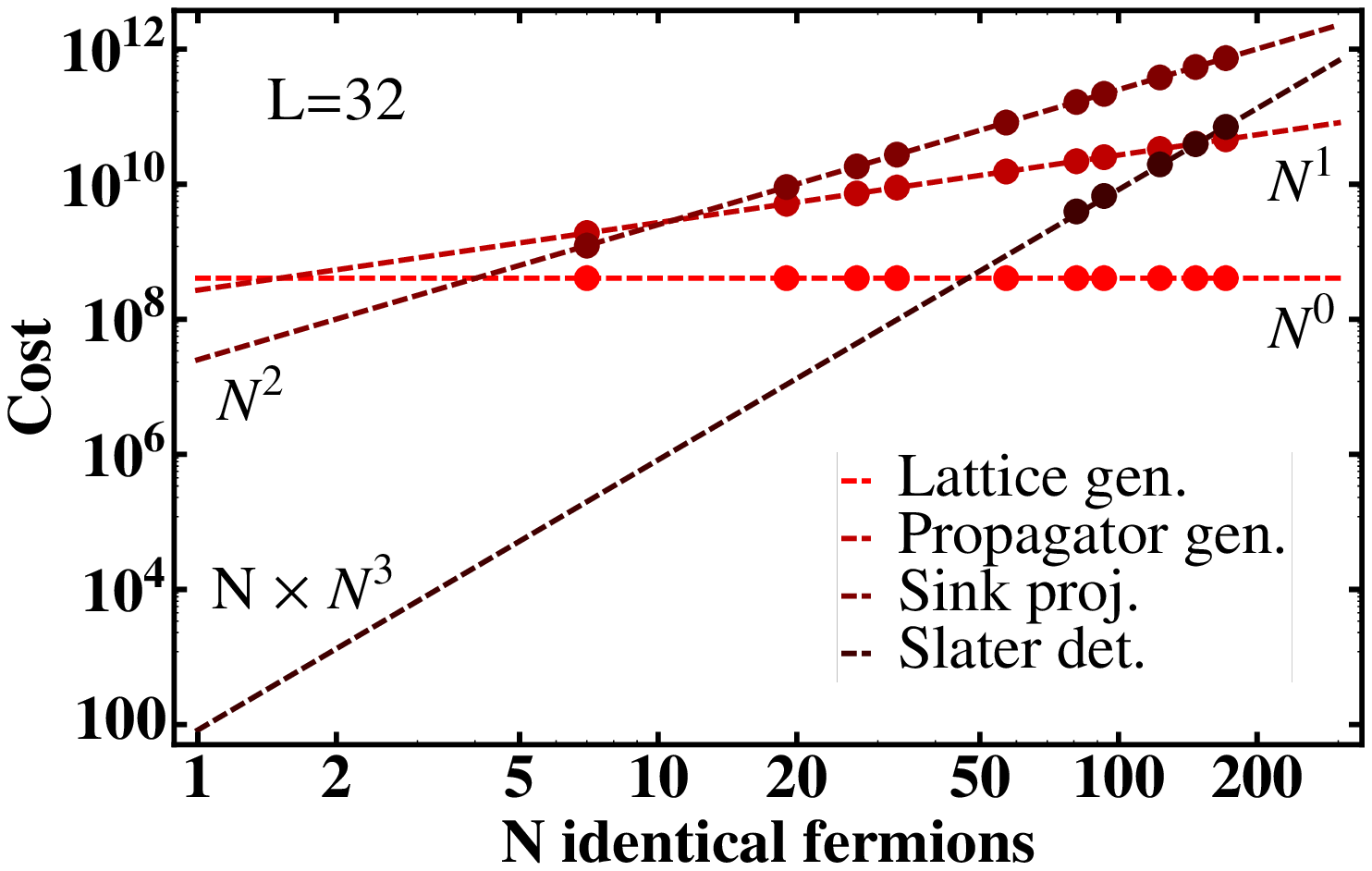}\caption{Cost of numerical simulations as a function of the number of fermions N.
}
\label{fig:cost}
\end{minipage}
\hspace{8pt}
\begin{minipage}[t]{0.486\textwidth}
\centering
\includegraphics[width=0.95\textwidth]{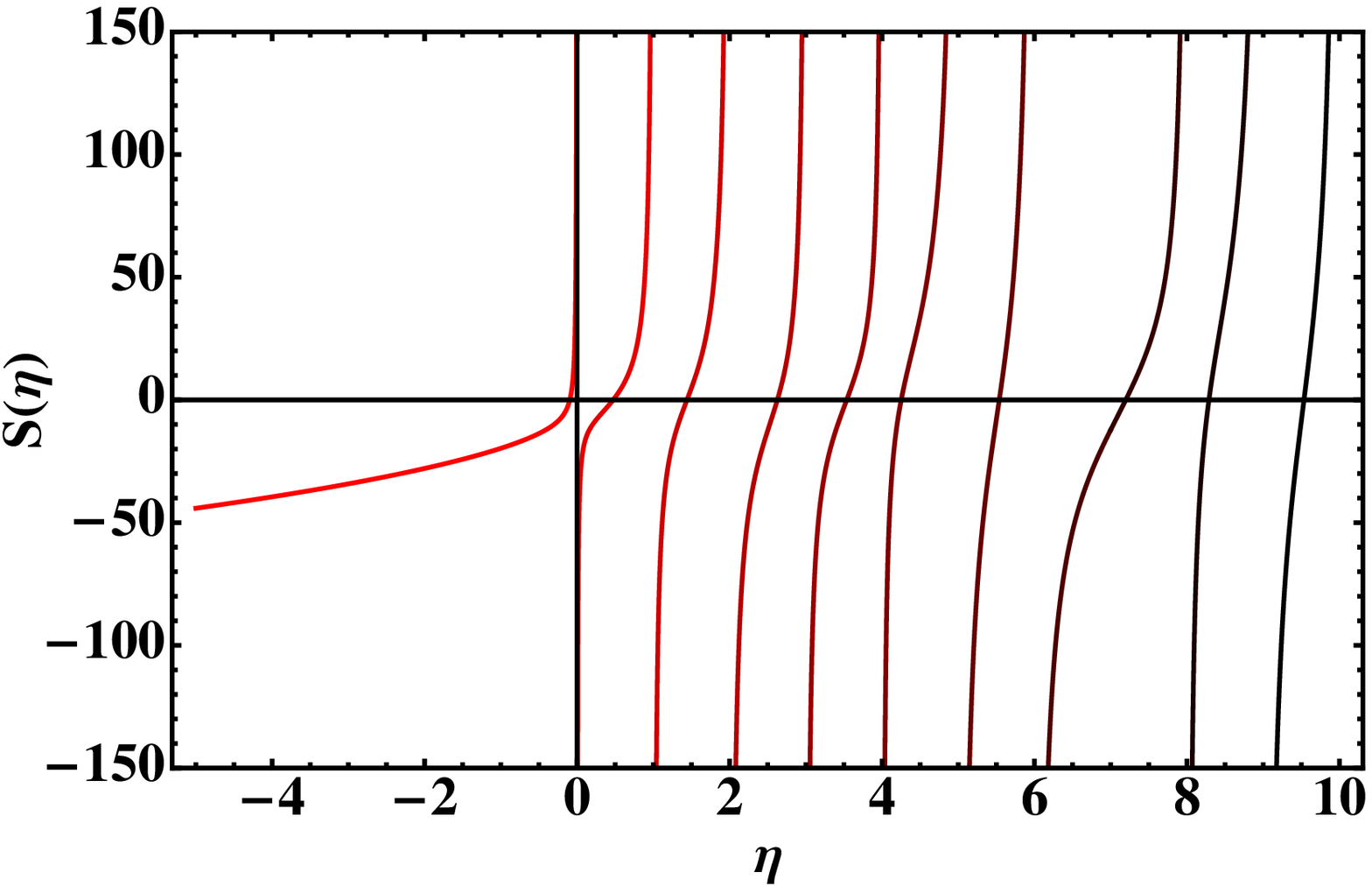}
\caption{A plot of the three-dimensional $\zeta$ function $S(\eta)$ given by Eq.~5.2}
\label{fig:s_plot}
\end{minipage}
\end{figure}

Numerical simulation of \Eq{action} consists of four steps: 1) lattice generation, 2) propagator generation, 3) projection of time-evolved states onto sinks and 4) computation of Slater determinants (i.e., anti-symmetrization of initial and final states).
The relative computational cost of each of these steps on an $L=32$ lattice is shown in \Fig{cost} as a function of the number of identical fermions.
To give meaning to the vertical axis, note that the $\calO(N^0)$ curve indicates the numerical cost of generating $\calO(L^3)$ random numbers (i.e., lattice generation).
The steps 1)-3) scale like $T\times L^3$ or $T\times L^3 \log{L^3}$, whereas step 4) is independent of the spatial volume.
Although step 4) scales like $\calO(N^4)$,\footnote{The cost of computing determinants scales like $N^3$, however, we compute determinants of all $N$ sub-matrices as well, giving rise to an additional power of $N$.} the computational cost of this step is negligible even for as many as $N\sim50-100$ identical fermions on the typical volumes we consider, which range from $L\sim12-64$.

\section{Transfer matrices}
\label{sec:4}

We may translate our lattice action \Eq{action} into Hamiltonian language with a suitable reinterpretation of the expression for multi-fermion correlation functions.
Such correlation functions are obtained from an ensemble average of direct products of propagators.
Since the propagator defined in \Eq{prop} is itself a product of uncorrelated random matrices (the auxiliary field is action-less), the multi-fermion correlation function will factorize into a matrix product of ensemble averages.
If we define:
\begin{eqnarray}
\calT = {\calD}^{1/2} (1-\calV) {\calD}^{1/2}\ ,
\end{eqnarray}
where
\begin{eqnarray}
\calD = \underbrace{ D^{-1/2} \otimes \ldots \otimes D^{-1/2} }_N  \ ,\qquad (1-\calV) = \underbrace{ \langle  X(\tau) \otimes \ldots \otimes X(\tau)\rangle }_N 
\end{eqnarray}
are $V^N$ dimensional matrices, then the N-fermion correlator may be written as:
\begin{eqnarray}
\underbrace{ \langle  K^{-1}(\tau;0) \otimes \ldots \otimes K^{-1}(\tau,0)\rangle }_N   = \calD^{-1/2} \calT^\tau \calD^{-1/2}\ ,
\end{eqnarray}
and we may identify $\calT$ as the transfer matrix and $\calH=-\log \calT$ as the Hamiltonian of the N-fermion system.

In the case of two fermions (one up and one down), the transfer matrix evaluated in momentum space is given by:\footnote{We have recently implemented a new Hermitian, Galilean invariant and analytic version of this interaction which corresponds to replacing: $ \sqrt{C(\bfp')} \sqrt{C(\bfq')} \rightarrow C(\bfp-\bfp')$; results using this improved interaction will be presented in detail in a future publication.}
\begin{eqnarray}
\langle \bfp' \bfq' | \calT | \bfp \bfq \rangle = \frac{ \delta_{\bfp,\bfp'} \delta_{\bfq,\bfq'} + \sqrt{C(\bfp')} \sqrt{C(\bfq')} \delta_{\bfp+\bfq,\bfp'+\bfq'}  }{ e^{-(\bfp^2+{\bfp'}^2 + \bfq^2 + {\bfq'}^2)/(4M)}}\ ,\quad  C(\bfp) = \frac{4\pi}{M} \sum_n C_{2n} O_{2n}(\bfp)\ ,
\end{eqnarray}
where we have expanded $C(\bfp)$ in the operator basis:
\begin{eqnarray}
O_{2n}(\bfp) = M^n \left(1-e^{-\bfp^2/M}\right)^n \approx \bfp^{2n}\ ,\quad \textrm{for $|\bfp|<<1$}\ .
\label{eq:ops}
\end{eqnarray}
This transfer matrix can be diagonalized exactly, with the zero center of momentum energy eigenvalues ($e^-{E}$) given by solutions to the integral equation:
\begin{eqnarray}
\sum_n C_{2n} I_{2n}(p) = 1\ ,\quad I_{2n}(p) = \frac{1}{V} \sum_{\bfq\in BZ} \frac{O_{2n}(\bfq)}{e^{(-p^2+\bfq^2)/M}-1}\ ,
\label{eq:lat_eigs}
\end{eqnarray}
where $p=\sqrt{M E}$ and $\bfq$ are momenta within the first Brillouin zone (BZ).

\section{Lattice parameter tuning and the continuum limit}
\label{sec:5}

In the continuum, Luscher's formula allows us to relate the two particle scattering phase shifts at infinite volume to the discrete energies of the same system at finite volume \cite{Luscher:1986pf,Luscher:1990ux}.
Having solved the two body problem exactly on the lattice at finite volume, we may now relate the lattice couplings of our theory to continuum scattering data.
Starting from the effective range expansion
\begin{eqnarray}
p\cot\delta_0 = -\frac{1}{a} + \frac{1}{2} r_0 p^2 + \ldots
\end{eqnarray}
one can relate the scattering length ($a$), effective range ($r_0$) and higher order shape parameters to the s-wave scattering phase shift.
Given an expression for $p\cot\delta_0$ one may then determine the continuum energy eigenvalues in a finite box from the solutions to \cite{Beane:2003da}:
\begin{eqnarray}
p \cot \delta_0 = \frac{1}{\pi L} S(\eta) \ ,\quad S(\eta) = \lim_{\Lambda\to\infty} \left[ \sum_{|\bfn|<\Lambda} \frac{1}{\bfn^2-\eta} - 4\pi\Lambda \right]\ ,
\label{eq:Sfunc}
\end{eqnarray}
with $\eta = (pL/2\pi)^2$.
Finally one may tune the lattice couplings $C_{2n}$ (for $n=1,\ldots,k$) defined in \Eq{ops} by matching the lattice eigenvalues predicted by \Eq{lat_eigs} to the lowest $k$ continuum energies predicted by Luscher's formula.
In this way, we may absorb all temporal and spatial lattice discretization errors into our definition of the couplings.

The continuum limit for our lattice theory corresponds to the limit of infinite scattering length as measured in units of the lattice spacing.
In order to maintain a finite physical scattering length, one must take this limit while keeping other physical quantities measured in units of the scattering length held fixed.
In the case of unitary fermions, however, we need not concern ourselves with such complications and simply take $p\cot \delta_0 =0$.
We therefore tune the couplings $C_{2n}$ so that the lattice eigenvalues match the lowest $k$ roots of $S(\eta)$ shown in \Fig{s_plot}.

\begin{figure}
\begin{minipage}[t]{0.486\textwidth}
\centering
\includegraphics[width=\textwidth]{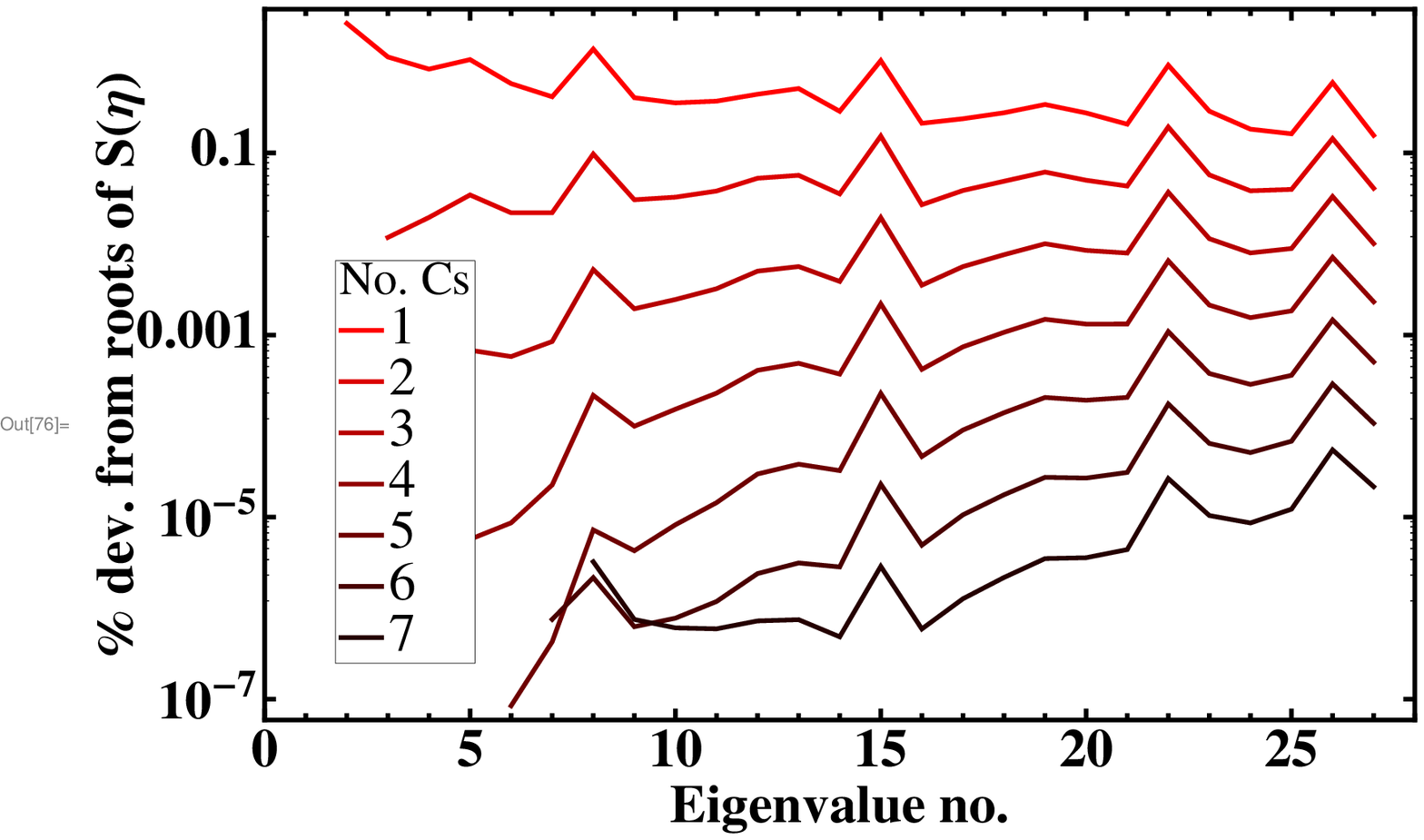}
\caption{Percent deviation in $\eta$ between exact lattice eigenvalues (using $M=5$ and $L=32$) and continuum Luscher eigenvalues.}
\label{fig:deviation}
\end{minipage}
\hspace{8pt}
\begin{minipage}[t]{0.486\textwidth}
\centering
\includegraphics[width=0.95\textwidth]{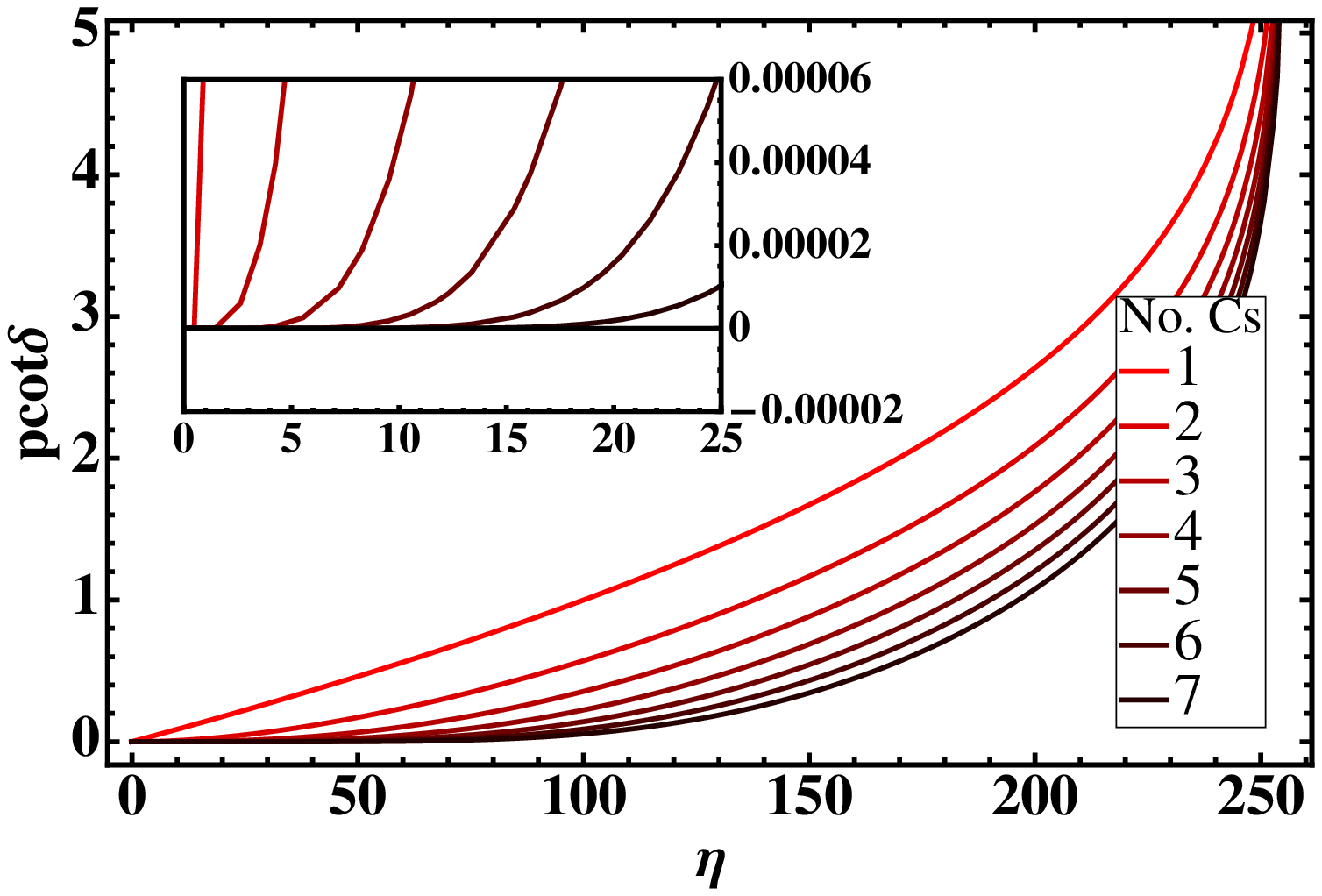}
\caption{Implied $p\cot\delta_0$ obtained from exact lattice eigenvalues (using $M=5$ and $L=32$)  and Luscher's formula.}
\label{fig:pcotdelta}
\end{minipage}
\end{figure}

In \Fig{deviation} we plot the percent deviation between the roots of $S(\eta)$ and the lattice eigenvalues predicted by \Eq{lat_eigs} for up to seven tuned $C$-values for unitary fermions.
For eigenvalues less than the $k$-th, the deviation is exactly zero by construction.
For eigenvalues beyond the $k$-th, we find that the percent deviation remains quite small even for eigenvalues as high as 27, corresponding to three filled shells.
Given a tuned set of $C$-values, we may also take the eigenvalues predicted by \Eq{lat_eigs} and insert them back into Luscher's formula, giving a lattice prediction for $p\cot \delta_0$.
\Fig{pcotdelta} shows a plot of the predicted $p\cot\delta_0$ for up to seven tuned $C$-values.
We find $p\cot\delta_0 \ll 1$ for a wide range of momenta, extending well beyond that of the $k$-th eigenvalue we tuned to.

\section{Conclusion}
\label{sec:6}

We have developed a new approach for simulating a large number of nonrelativistic fermions in the unitary regime.
In these proceedings we have described some of the details of our lattice construction, an efficient numerical implementation of the theory and a method for tuning the lattice couplings to scattering data.
Application of these ideas to unitary fermions in a finite box \cite{Lee:2010qp} and in a harmonic trap\footnote{Details regarding the introduction of external potentials will be presented in \cite{Nicholson:2010ms}.} \cite{Nicholson:2010ms} are discussed in greater detail in our companion proceedings.
There the issue of finding optimal sinks/sources as well as finite volume and cutoff effects are explored, and preliminary results for the Bertsch parameter and pairing gap reported.

\section{Acknowledgments}
\label{sec:7}
This work was supported by U. S. Department of Energy grants DE-FG02-92ER40699 (to M. G. E.) and DE-FG02-00ER41132 (to D. B. K., J-W. L. and A. N. N.).
M. G. E is supported by the Foreign Postdoctoral Researcher program at RIKEN.

\bibliography{lattice2010}

\begin{thebibliography}{10}

\bibitem{Lepage:1989hd}
G.~P. Lepage,
\newblock Invited lectures given at TASI'89 Summer School, Boulder, CO, Jun
  4-30, 1989.

\bibitem{2008AnPhy.323.2952T}
S.~{Tan},
\newblock Annals of Physics {\bf 323}, 2952 (2008), cond-mat/0505200.

\bibitem{2008AnPhy.323.2971T}
S.~{Tan},
\newblock Annals of Physics {\bf 323}, 2971 (2008), cond-mat/0508320.

\bibitem{2008AnPhy.323.2987T}
S.~{Tan},
\newblock Annals of Physics {\bf 323}, 2987 (2008), arXiv:0803.0841.

\bibitem{Chang:2004zz}
S.~Y. Chang, V.~R. Pandharipande, J.~Carlson, and K.~E. Schmidt,
\newblock Phys. Rev. {\bf A70}, 043602 (2004), physics/0404115.

\bibitem{Chang:2007zz}
S.~Y. Chang and G.~F. Bertsch,
\newblock Phys. Rev. {\bf A76}, 021603 (2007), physics/0703190.

\bibitem{2010arXiv1008.3191B}
D.~Blume and K.~M. Daily,
\newblock (2010), arXiv:1008.3191.

\bibitem{Lee:2010qp}
J.-W. Lee, M.~G. Endres, D.~B. Kaplan, and A.~N. Nicholson,
\newblock (2010), arXiv:1011.3026.

\bibitem{Nicholson:2010ms}
A.~N. Nicholson, M.~G. Endres, D.~B. Kaplan, and J.-W. Lee,
\newblock (2010), arXiv:1011.2804.

\bibitem{Chen:2003vy}
J.-W. Chen and D.~B. Kaplan,
\newblock Phys. Rev. Lett. {\bf 92}, 257002 (2004), hep-lat/0308016.

\bibitem{Luscher:1986pf}
M.~Luscher,
\newblock Commun. Math. Phys. {\bf 105}, 153 (1986).

\bibitem{Luscher:1990ux}
M.~Luscher,
\newblock Nucl. Phys. {\bf B354}, 531 (1991).

\bibitem{Beane:2003da}
S.~R. Beane, P.~F. Bedaque, A.~Parreno, and M.~J. Savage,
\newblock Phys. Lett. {\bf B585}, 106 (2004), hep-lat/0312004.

\end{thebibliography}
\bibliographystyle{h-physrev.bst}

\end{document}